# Neonatal EEG Interpretation and Decision Support Framework for Mobile Platforms

Mark O'Sullivan, Sergi Gomez, Alison O'Shea, Eduard Salgado, Kevin Huillca,
Sean Mathieson, Geraldine Boylan, Emanuel Popovici, Andriy Temko

*Abstract*— **This paper proposes and implements an intuitive and pervasive solution for neonatal EEG monitoring assisted by sonification and deep learning AI that provides information about neonatal brain health to all neonatal healthcare professionals, particularly those without EEG interpretation expertise. The system aims to increase the demographic of clinicians capable of diagnosing abnormalities in neonatal EEG. The proposed system uses a low-cost and low-power EEG acquisition system. An Android app provides single-channel EEG visualization, traffic-light indication of the presence of neonatal seizures provided by a trained, deep convolutional neural network and an algorithm for EEG sonification, designed to facilitate the perception of changes in EEG morphology specific to neonatal seizures. The multifaceted EEG interpretation framework is presented and the implemented mobile platform architecture is analyzed with respect to its power consumption and accuracy.**

## I. Introduction

Electroencephalography (EEG) is the most common tool available for the real-time assessment of neonatal brain health and for the detection of abnormalities such as seizures. However, the acquisition and interpretation of neonatal EEG can be a difficult task for non-experts and it is challenging to implement in the acute setting. Failure to detect and treat abnormal activity can lead to life-threatening outcomes [1].

Despite significant research being conducted in the area of portable electronics, sensors and signal processing, truly portable EEG monitoring is not yet widely used in the clinical setting [2]. The gold standard, multichannel EEG, requires expensive acquisition systems (≈ €25,000) and significant clinical expertise, which is not widely available. This often entails lengthy patient preparation, such as skin impedance reduction with abrasive cream and improvement of conductivity using conductive gel. Few staff in the neonatal intensive care unit (NICU) have received sufficient training to interpret neonatal multichannel EEG. As a compromise, a simpler form of EEG called amplitude integrated EEG, or aEEG is often used. Although aEEG certainly improves the situation in units with no other means of neurological monitoring, multiple studies have shown that aEEG has several limitations and is still overly reliant on staff who are experts in interpretation of the aEEG [3]. Therefore, significant research has been conducted in the space of artificial intelligence (AI) algorithms that are able to alert NICU staff about abnormal brain functioning such as seizures [1] or abnormal EEG background activity [4]. These systems are designed to provide decision support to the clinician regarding the diagnosis and treatment of such events. Their transition to clinical use has improved in recent years, but a suitable interface for interaction between the decision support tool and clinic staff is still required. The 'black-box' element of the AI interpretation can be moderated through the use of alternative methods such as sound-based EEG interpretation. Human hearing is superior to the visual sense at assessing both the spatial and temporal evolution of signal frequency characteristics [5]. Hearing allows for faster processing and releases visual sense for other tasks [6]. To address the identified issues such as cost, portability and expert availability, we have previously presented a novel sound-based interpretation algorithm [7].

This paper presents a single channel, handheld, mobile, dry EEG monitoring system suitable for use in acute and resource-limited settings. This technology can offer virtually instantaneous monitoring of the neonatal brain health to all units, particularly those that currently have no objective means of neurological monitoring. The system combines the aforementioned approaches to facilitate the acquisition and non-expert interpretation of neonatal EEG through an exemplar application such as seizure detection. The overview of the system is depicted in Fig. 1. A low-cost, portable EEG acquisition board (€200) was used with dry EEG electrodes. The developed app provides means for subjective and objective assessment of the neonatal brain health, such as EEG visualization, traffic-light indication of neonatal seizures facilitated by a convolutional neural network (CNN) and sound-based EEG interpretation, which was tuned to emphasize changes in EEG morphology specific to neonatal seizures. The system was analyzed from performance, usability and power consumption perspectives.

## II. Methodology

### A. Hardware: EEG Sensors and Acquisition System

In order to capture the low-amplitude (±100μV) EEG potentials from the scalp, strenuous skin preparation is often required. This is usually accomplished using wet electrodes, which entails skin abrasion and the application of conductive gels, requiring clinical expertise and vital time.

Research was supported in part by Wellcome Trust Seed Award (200704/04/Z/16/Z), SFI INFANT Centre (12/RC/2272) and TIDA (17/TIDA/504) and HRB (KEDS-2017-020).

M. O'Sullivan, A. O'Shea, E. Salgado, S. Gomez, K. Huillca, S. Mathieson, G. Boylan, A. Temko are with Irish Centre for Fetal and Neonatal Translational Research (INFANT), University College Cork.
E. Popovici is with the School of Engineering, UCC.
(email: mark.e.osullivan@umail.ucc.ie)

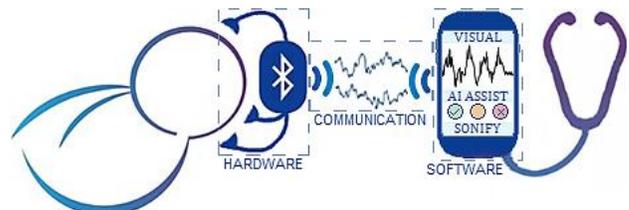

Figure 1. Proposed EEG acquisition & interpreation system

TABLE I.  EEG INSTRUMENTATION & VISUALIZATION STANDARDS

|      | Fs (Hz) | Res (bits) | Rin (MΩ) | CMRR (dB) |
|------|---------|------------|----------|-----------|
| IFCN | 200     | 12         | 100      | 110       |

|      | Datarate     | Horizontal | Vertical    | Resolution |
|------|--------------|------------|-------------|------------|
| IFCN | 120 sample/s | 30 mm/s    | 10 mm/chan  | 4 pix/mm   |

The acquisition electronics must be precise, as per the International Federation of Clinical Neurophysiology (IFCN) guidelines in Table I [8]. OpenBCI Ganglion bio-sensing device encapsulates high-precision electronics, in a low-cost, battery-powered form factor. The board can acquire 4 channels at 24bit resolution and 200Hz sampling rate. The data is Delta Encoded, resulting in a 4kB/s data-rate. Bluetooth Low Energy (BLE) via a Simblee Radio Module enables wireless transfer the data from the board to the app.

The proposed single channel, dry electrode system allows for easy movement to various locations on the head to provide a thorough examination for abnormal activity. Previous studies have explored the use of dry electrodes for EEG, which rely on novel mechanical designs to acquire accurate EEG signals without significant skin preparation [9]. The experiments conducted in [10] indicated that large impedances associated with dry electrodes do not corrupt the EEG signals enough to significantly affect signal accuracy. The EEG acquisition board achieved over 0.99 correlation with clinical grade EEG.

### B. Software: Pre-processing and EEG Visualization

In order to improve portability, an Android application was developed and deployed on a Samsung A6 tablet (2GB RAM, 8 cores @ 1.6GHz CPU, 7300mAh battery). The app performs BLE pairing, decoding, parsing and storage of the data. Using the Battery Historian developed by Google, the tablet's RAM and power consumption, and CPU usage in percentage was monitored over the course of twelve 1-minute executions of the app [11]. The mean and standard deviation (SD) values for each of the metrics were computed.

Visual representation of EEG was implemented on the tablet according to IFCN guidelines as in Table I. Horizontal screen display scaling of 1 second to 30mm, vertical scaling of 10mm per channel, with a minimum display resolution of 120 data points per second is advised. Due to the complexity of EEG, digital filtering is essential for precise interpretation. High and low-pass filters with variable cut-off frequencies and a 50Hz notch filter were implemented on the app. A single channel of EEG was plotted in real-time. The RAM usage was separately monitored during the plotting functions.

### C. Sound-Based Subjective Interpretation

The aim of neonatal EEG sonification is to increase the demographic of clinical staff capable of understanding and accurately assessing EEG through the development of a more intuitive method to subjectively interpret the EEG. Seizure EEG presents rhythmically evolving low-frequency activity and methods that can highlight this evolution in the audio domain may provide a new means of interrogating neonatal brain health. In this work, two methods were developed and compared with visual EEG assessment by non-EEG experts in terms of their sensitivity to detection of neonatal seizures. The core concepts of the sound-based interpretation algorithms are presented.

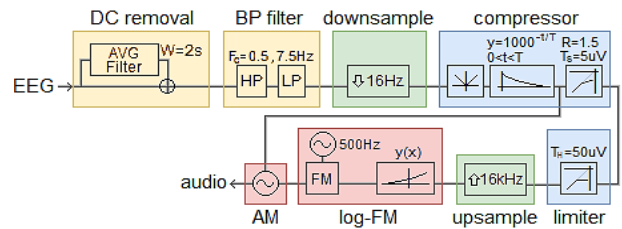

Figure 2. Audio Synthesis (FM/AM) Process.

#### 1) Phase Vocoder

The phase vocoder (PV) algorithm, which was previously developed in [15], transforms the low frequency EEG signal (0.5 - 16Hz) into the audio frequency range (500 - 16kHz) by preserving the spectral envelope of the original EEG signal and keeping horizontal phase coherence within a given frequency bin. Using the PV, the signal can be time compressed to rapidly review a large amount of EEG data. The time-scale factor impacts the resulting audio features.

#### 2) Audio Synthesis

Frequency modulation (FM) synthesis is presented in Fig. 2. It consists of varying the frequency of a carrier wave with a modulating wave. The amount the carrier varies around its average is proportional to the amplitude of the modulating wave [16]. This methodology was adopted for use with EEG. A sinusoidal carrier frequency of 500Hz is modulated with the time content of the EEG signal. Therefore, EEG signal oscillations drive the frequency of the resulting audio. The amplitude and rhythmic activity of the EEG contains equally vital information. Thus, amplitude modulation (AM) was used to embed the amplitude characteristics of the EEG in the audio.

The data is filtered between 0.5-7.5Hz and down-sampled to 16Hz. An amplitude compressor is applied, which reduces the dynamic range of the signal with minimal spectral distortion. The envelope of the signal is extracted and used to reduce the amplitude of the signal proportional to a given ratio (R=1.5), when a given threshold (T=5µV) is exceeded. A limiter is used to clip the amplitude to 50µV to avoid aliasing. FM is applied to map the voltage to frequency, as determined by (1), generating a signal in the 50-5kHz range, centered at 500Hz. This range was chosen as it encompasses the most sensitive region of human hearing system. Finally, the amplitude envelope of the EEG signal is used to vary the amplitude of the audio signal through AM synthesis.

$$y(x) = \frac{2f_L}{F_s}\left(\frac{f_H}{f_L}\right)^{\frac{x/T_H+1}{2}} \quad (1)$$

where $f_S$=200Hz, $f_L$=50Hz, $f_H$=5kHz, and $T_H$=50µV.

In contrast to PV, FM supports the perception of evolution of frequency in time and also the presence of rhythm, which is evident when repetitive structures emerge in the EEG signal.

#### 3) Survey Design

In order to optimize and quantitatively assess the performance of the sonification algorithms, a survey was conducted amongst a cohort (N = 11) of assessors who were not EEG experts. An anonymized database of neonatal seizure and background activity was created by an expert neonatal neurophysiologist (GB) who annotated over 1300 seizures. Using the seizure detection algorithm from [7], per-channel annotations were derived from overall seizure annotations using the maximum probability across EEG channels. From this, a new database was compiled with the help of a neonatal neurophysiologist (SM), which consisted of typical seizure and non-seizure activities (including seizure-like artifacts such as respiration, pulsatile, sweat and bad-electrode). In total, 100 single channel seizures and 100 background activity segments were selected from a variety of subjects, channels, with varying durations and morphologies.

TABLE II. COMPUTATIONAL COST RESULTS

|  | CPU (%) | RAM (MB) | Power (mAh) |
|---|---|---|---|
| Communication | 10.72±1.16 | 25.34±0.38 | 19.13 |
| Visualization | ~ | 43.64±0.52 | ~ |
| Phase Vocoder | 38.00±04 | 10.00±03 | 10.4 |
| CNN 6 | 21.29±2.8 | 23.33±1.18 | 30.3 |
| CNN 11 | 25.63±1.0 | 25.96±0.72 | 38.21 |

TABLE III. CNN PERFORMANCE RESULTS

| layers | no. params | AUC (%) | AUC90 (%) | GDR(0.5fd/h) (%) |
|---|---|---|---|---|
| 6 | 17058 | 97.1 | 82.9 | 78.8 |
| 11 | 286422 | 97.7 | 86.5 | 83.2 |

The survey assessed the detection rate using: 1) visual EEG traces 2) PV with a time-scale factor of x1, x5 and x10; 3) FM/AM with time-scale factor of x1, x5, x10. Training data (3 seizure and 3 non-seizure cases) was made available for each method to develop the participants understanding of the major differentiating characteristics of seizure sounds with respect to the non-seizure sounds. The user then assessed 10 randomly selected examples from the corpus. The average and individual seizure detection accuracies and 95% confidence intervals (CI) were computed.

*D. AI-Assisted Objective Interpretation*

A deep convolutional neural network (CNN) was previously developed for the task of neonatal seizure detection [12]. It outperformed the previous state-of-the-art Support Vector Machine (SVM) based system. Convolutional filters act as data-driven feature extractors, which hierarchically extract features of increasing complexity to give maximum class discrimination, without any dense (fully connected) layers. The CNN architecture enables having the feature extraction and classification stages in one routine for end-to-end system optimization. Over 900 hours of multi-channel full-term EEG was used to test the algorithm [12].

The main advantage of the CNN-based detector is the usage of raw EEG as an input to the system. This can be contrasted to a set of complex hand-crafted features, as used in SVM-based systems [13]. SVM would require more time and computational expense to implement and run. Two CNN architectures were implemented on Android with 6 and 11 layer depths (the number of convolutional layers). The 11-layer architecture allows for more complex feature extraction but also increases processing time and power consumption. Performance comparison of the 6 and 11 layer algorithms is reported in [14]. Here, the two architectures are contrasted with more clinically relevant event-based metrics such as good detection rate (GDR) and the number of false detections per hour (FD/h). The power consumption of the forward pass of the 6 and 11 layer CNN was computed and the trade-off between improved accuracy of a deeper architecture with respect to its computational cost is discussed.

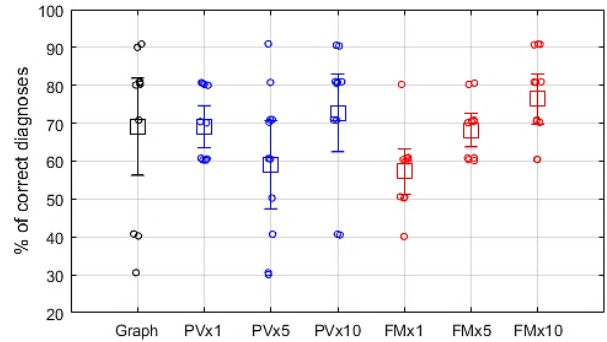

Figure 3. EEG interpretation method survey results, with the mean (□), 95% CI and individual participant (o) scores indicated.

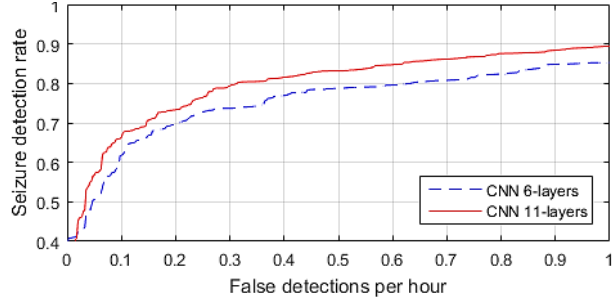

Figure 4. CNN 6 vs. 11 layer seizure detection rate.

III. RESULTS & DISCUSSION

An Agilent N6705A DC power analyzer was used to monitor the current draw of the OpenBCI board. The average current draws during idle and streaming modes were 14.2±0.19mA and 15.5±0.16mA, respectively. These values match the values presented in the product datasheet [17]. At this rate, continuous streaming of EEG could be performed for roughly 130 hours using four AA batteries (2000mAh).

Table II displays the computational expense results from Battery Historian. Visualization is an intrinsic part of the functionality of the mobile devices and does not consume additional power. The acquisition (BLE connection, parsing, storing) elements are notably inexpensive, with CPU of 10.72% and power consumption of 19.13mAh. This is due in part to the use of BLE, which consumes 10 times less power than standard Bluetooth. However, this comes at a cost of lower connectivity range and data-throughput [18].

Table II also presents the results that were previously reported in [19] for the PV algorithm implemented on an Android device. It is worth noting that a lower quality Android device was used to obtain these values and they can serve as the maximum system requirements needed to facilitate the entire acquisition and multifaceted interpretation system. The previously developed PV (x1) algorithm was thoroughly optimized to perform across multiple threads. Although the algorithm performs the difficult task of interpolating the EEG signal from 16Hz to 16kHz, the CPU remains below 40% even on a lower quality device with 1GB Ram and 4core 1.25GHz CPU. In total, it can be seen that a total of 98mAh and 128MB of RAM is consumed by the app. The total CPU usage is estimated to be approximately 95%. Through efficient optimization of the app this can be significantly reduced [19].

Fig. 3 presents the results of the subjective interpretation survey. Among the real-time methods (graphical, PVx1 and FMx1), the PVx1 method achieves the highest percentage of correctly diagnoses at 69%, while maintaining the narrowest CI. It can be seen that the perception of time-evolving seizure morphologies seems to be facilitated by speeding up the audio domain signals. PV and FM at high speeds (x10) achieve the highest results of 73% and 76% respectively. Interestingly, the graphical interpretation achieves the widest CI, implying that non-expect visual interpretation is very inconsistent. These preliminary results warrant further analysis into the complementarity of sound-based and graphical interpretation. An improved sonification method combining FM and PV, enabling the perception of rhythm and frequency evolution is currently under development.

The performance of an AI-assisted interpretation for two CNN architectures is presented in Fig. 4. It can be seen that the deeper 11-layer system consistently outperforms the 6-layer alternative for the entire range of FDs/h [14]. The computational cost results presented in Table II indicate that the CPU, RAM and battery consumptions are larger for the deeper 11-layer architecture with the power consumption of the 11-layer architecture increased by 7.9mAh. This will result in the battery lasting approximately 19 minutes less with 11 layers than with 6, given that when idle, the tablet only lasts ~18 hours. As the application is required to run in real-time, the processing speed is of interest. In order to compute an 8 second window with a 1 second shift, the 6-layer architecture takes 115ms. In contrast, 283ms are required for the 11-layer one. This reduction in computational speed may cause a computational bottleneck once acquisition, visualization and sonification are all active.

IV. CONCLUSION

The proposed acquisition and multifaceted (objective and subjective) interpretation system was successfully implemented on low-cost electronics and an Android device. The total power consumption of the Android app is roughly 100mAh, with an average RAM consumption of 128MB. The battery life allows for long-term continuous EEG monitoring without charging. It is shown that clinical performance of the CNN algorithm increases significantly with the convolutional depth. However, the additional computational complexity causes an increase in power consumption, and more significantly, processing time.

The survey results show that EEG sonification is a valid alternative method for assessing seizure activity by non-experts. Lower variance is achieved using sonification methods and an increase in the percentage of correct diagnoses is observed. These results will provide the pathway for further refining the developed algorithms and conducting larger cohort studies. While the current setup targets neonatal seizures, the capabilities of the system are extendible to facilitate quick screening for asphyxia induced abnormalities in background neonatal EEG acquired soon after birth. Given the low cost (€200) of the device and pervasive nature of the interpretation framework, the proposed system may bridge the gap between tertiary care and primary/low-resource settings by providing intermediate assessment of neonatal brain health.